# The discriminant power of RNA features for pre-miRNA recognition

Ivani de O. N. Lopes[*,1,2,3] and Alexander Schliep[2] and André C. P. de L. F. de Carvalho[3]

[1]Empresa Brasileira de Pesquisa Agropecuária, Embrapa Soja,
Caixa Postal 231, Londrina-PR, CEP 86001-970, Brasil
[2]Department of Computer Science and BioMaPs Institute for Quantitative Biology
Rutgers University
110 Frelinghuysen Road, Piscataway, NJ, 08854, USA
[3]Instituto de Ciências Matemáticas e de Computação
Avenida Trabalhador são-carlense, 400 - Centro, São Carlos - SP, Brasil

Email: Ivani de O. N. Lopes*- ivani.negrao@embrapa.br; Alexander Schliep - schliep@cs.rutgers.edu; André C. P. de L. F. de Carvalho - andre@icmc.usp.br;

[*]Corresponding author

## Abstract

**Background**: Computational discovery of microRNAs (miRNA) is based on pre-determined sets of features from miRNA precursors (pre-miRNA). These feature sets used by current tools for pre-miRNA recognition differ in construction and dimension. Some feature sets are composed of sequence-structure patterns commonly found in pre-miRNAs, while others are a combination of more sophisticated RNA features. Current tools achieve similar predictive performance even though the feature sets used - and their computational cost - differ widely. In this work, we analyze the discriminant power of seven feature sets, which are used in six pre-miRNA prediction tools. The analysis is based on the classification performance achieved with these feature sets for the training algorithms used in these tools. We also evaluate feature discrimination through the F-score and feature importance in the induction of random forests.

**Results**: More diverse feature sets produce classifiers with significantly higher classification performance compared to feature sets composed only of sequence-structure patterns. However, small or non-significant differences were found among the estimated classification performances of classifiers induced using sets with diversification of features, despite the wide differences in their dimension. Based on these results, we applied a feature selection method to reduce the computational cost of computing the feature set, while maintaining discriminant power.

We obtained a lower-dimensional feature set, which achieved a sensitivity of 90% and a specificity of 95%. Our feature set achieves a sensitivity and specificity within 0.1% of the maximal values obtained with any feature set (SELECT, Section 2) while it is 34 times faster to compute. Even compared to another feature set (FS2, see Section 2), which is the computationally least expensive feature set of those from the literature which perform within 0.1% of the maximal values, it is 34 times faster to compute. The results obtained by the tools used as references in the experiments carried out showed that five out of these six tools have lower sensitivity or specificity.

**Conclusion**: In miRNA discovery the number of putative miRNA loci is in the order of millions. Analysis of putative pre-miRNAs using a computationally expensive feature set would be wasteful or even unfeasible for large genomes. Comprising even false positive rates or accuracy as low as 5% is not an option, as would lead to hundreds of thousands of additional pre-miRNA candidates for verification. Consequently, to make the analysis of putative miRNA using ab-initio tools computationally feaseable, new tools with low computational cost and high predictive performance are needed. In this work, we propose a relatively inexpensive feature set and explore most of the learning aspects implemented in current ab-initio pre-miRNA prediction tools, which may lead to the development of efficient ab-initio pre-miRNA discovery tools.

The material to reproduce the main results from this paper can be downloaded from http://bioinformatics.rutgers.edu/Static/Software/discriminant.tar.gz.

## Background

A **microRNA** (miRNA) is a small (approx. 17-25 nucleotides) non-coding RNA molecule (ncRNA) that modulates the stability of mRNA targets and their rate of translation into proteins [1]. MiRNAs are present in the genome of vertebrates, plants, algae and even viruses and are involved in diverse and complex biological processes, like development and cell differentiation [2], tumorigenesis [3] and immunity [4]. They can also alter plant gene expression in response to environmental stresses [5].

In animals, maturation of canonical miRNAs occurs in two steps: First, the long primary miRNA transcript is processed within the nucleus into a ∼60-120 nucleotides (nt) stem-loop hairpin precursor (pre-miRNA) by the enzyme Drosha [6]. Afterwards, within the cytoplasm, the enzyme Dicer cleaves the pre-miRNA into a double stranded RNA duplex (miRNA/miRNA*) and into a loop. The loop is degraded as a by-product [7], whereas the RNA duplex is unwound by helicase activity, releasing the mature miRNA and the star sequence [6]. The last is typically degraded whereas the mature miRNA guides the microribonucleoprotein complex (miRNP) to target messengers RNAs (mRNAs) by partial sequence complementarity [7].

**Machine learning in miRNA recognition**

Unlike protein-coding genes, ncRNA genes do not contain easily detectable signals [8]. Therefore, computational pipelines for miRNA discovery rely on characteristic features of pre-miRNAs. Coupled with comparative genomic computational pipelines, using RNAseq read libraries or completely ab-initio methods, machine learning (ML) algorithms have played an important role in miRNA discovery [7, 9–20]. ML algorithms induce models which are able to predict novel miRNAs based on patterns learned from known pre-miRNA sequences and from other RNA hairpin-like sequences, such as transfer RNA (tRNA) and mRNAs. ProMiR [9, 21], a probabilistic method, searches for pre-miRNA in genomic sequences using sequence and structure features. Naïve Bayes based probabilistic models were adopted to score miRNAs [11] and pre-miRNA [7, 19] candidates. Support vector machines (SVMs) [10, 12, 13, 16, 17], random forest (RF) [14, 20], relaxed variable kernel density estimator (RVKDE) [15] and generalized Gaussian components based density estimation ($G^2DE$) [18] were used to induce classifiers for pre-miRNA prediction.

**A limitation of working directly with the sequence** is that the information available may not be sufficient to infer accurate models for novel miRNA prediction. For example, the approximate number of RNAs folding into hairpin-like secondary structures in the human genome, without filtering based on phylogenetic conservation, was estimated as 11 million [22]. Filters derived from other pre-miRNA features reduced the number of pre-miRNA candidates to around 5,300. Therefore, the use of features that consider different aspects may reduce false positive rates in miRNA detection.

**Feature sets investigated in the literature**

Some of the features commonly extracted from RNA sequences for pre-miRNA recognition may not help to distinguish between positive (true pre-miRNAs) and negative (pseudo pre-miRNAs) classes. Therefore, the feature sets considered may have an important effect in the learning process. Several feature sets have been proposed for pre-miRNA recognition [10, 13, 14, 16, 18, 20, 23]. To comparatively evaluate the effect of these different sets, we investigated seven feature sets proposed in the literature, named here $FSi$, $i \in \{1,..,7\}$. They were used to induce six classifiers and they contain most of the features employed in computational pipelines for pre-miRNA discovery. Next, we briefly define each of these sets.

The first set, named FS1 [16], has 48 sequence and structural features. The second feature set FS2, which corresponds to a subset of 21 features of FS1, was used by [16] to induce a classifier, named microPred. The third feature set FS3, was used to induce a classifier called $G^2DE$ [18] and is composed by seven features also present in FS1. FS4 is a set of 32 sequence-structural features used by the triplet-SVM [10]. FS5 is a set of 1,300 sequence-structure motifs used by another classifier, mirident [23]. FS6 is the feature set used by the MiPred [14]. This feature set merged FS4, the minimum free energy of folding (MFE) and a stability measure (randfold). Finally, FS7 merged FS2 with four features of FS4 and three other features: percentage of low complexity regions detected in the sequence, maximal length of the amino acid string without stop codons and cumulative size of internal loops. It was used to induce the HuntMi [20]. The classification performances reported by the mentioned tools are among the highest in previous works. Their approximate specificities and sensitivities are: triplet-SVM (90%, 93%), MiPred (93%, 89%), microPred (97%, 90%), $G^2DE$ (98%, 87%), mirident (99%, 98%) and HuntMi (97%, 95%).

**Proposal and key findings**

In this study, we investigated the discriminative power of seven RNA feature sets, previously adopted in six tools developed for pre-miRNA prediction. Among them are two sets composed of sequence-structure features (FS4 and FS5) and five sets are a miscellany of RNA features (FS1-FS3 and FS6-FS7). The investigation of a specific feature set, using a particular training data and learning algorithm, may insert learning biases. As a consequence, the predictive performance for other training sets and learning algorithms could be different. To deal with this problem, we evaluated each feature set using the learning algorithms, SVMs, RF and $G^2DE$, which were used in the publications proposing those feature sets. According to the experimental results, the miscellaneous feature sets produced more accurate predictive models than features sets composed from only sequence-structure patterns. However, the differences in accuracy among miscellaneous feature sets are small or insignificant, despite their large differences in composition and dimensionality. Inspired by these results, we selected a subset of 13 features, of lower computational cost, but with a similar classification performance, when compared with FS1-FS3 and FS6-FS7 feature sets. The classes of positive and negative test sets used in the experiments presented in this paper were predicted by the tools that we used as reference. Except for one tool, higher sensitivity was tied to lower specificity and vice versa.

## 1 Material and Methods

Our goal was to investigate the predictive performance of RNA features in distinguishing pre-miRNAs from pseudo hairpins. As such, we adopted seven feature sets and three learning algorithms. The feature sets were used to induce classifiers for pre-miRNAs in triplet-SVM (FS4), MiPred (FS6), microPred (FS2 and FS1), $G^2DE$ (FS3), mirident (FS5) and HuntMi (FS7). SVMs were used in triplet-SVM, microPred and mirident, whereas RF was used in MiPred and HuntMi. Generalized Gaussian density estimator ($G^2DE$) [24] is not a tool for pre-miRNA prediction in the sense that the features have to be computed by the user in his/her own pipeline. Nevertheless, we included $G^2DE$ because of its predictive performance and class distribution interpretability. The subsections below provide details of our experiments.

*Data sets*

Human pre-miRNA sequences were downloaded from mirbase 19 [25] as the primary source of positive examples. In an attempt to avoid overfitting, we removed redundant sequences. As such, we clustered the available 1,600 human pre-miRNAs sequences using dnaclust [26] such that sequences within a cluster shared 80% similarity. Then, one sequence of each cluster was randomly picked. This yielded the set of positives composed of 1,377 non-redundant pre-miRNAs sequences.

The negative examples were the 8,494 pseudo hairpins from human RefSeq genes, originally obtained by [10] and subsequently used by [13, 14, 16, 18, 20, 23]. These sequences were obtained in order to keep basic features such as length distribution and minimal free energy of folding (MFE), similar to those observed in human pre-miRNAs. Moreover, this set has no redundant sequences. But, in order to adopt a uniform criterion for redundancy removal, we applied the same procedure adopted in the positive set. Only singleton clusters were formed. In practice, it is expected that high similarity between positive and negative examples leads to higher specificity [10].

*Experiments*

The predictive performance of any model is dependent on the training set representativeness, which usually increases with the increase in the training set size. Typically, density based algorithms, such as $G^2DE$, are more sensitive to the course of dimensionality and larger training sets are more likely to provide higher predictive performances. We determined experimentally the training set size which would be suitable for any algorithm and feature set. Each experiment was repeated 10 times, in order to provide standard deviations of each classification performance estimation. One repetition consisted of a test set, named here as GEN and 13 training sets. For a given repetition, GEN was composed by 459 sequences of each class, which corresponded to 1/3 of the 1,377 non-redundant pre-miRNA sequences. The remaining positive and negative sequences were used to sample increments of 67 sequences of each class to compose the training sets of $134, 268, ..., 1,742$ instances. Each feature set was computed from the same training and test sets. In total, we worked with 10 test sets and $13 \times 10$ training sets. The classification performances of the three algorithms converged to a threshold for training set sizes equal to 1,608, for all feature sets. Therefore, we presented the results for the largest training set, which contained 1,742 sequences.

*Classification performance measures*

Classification performance was measured as accuracy (Acc), sensitivity (Se), specificity (Sp), F-measure (Fm) and Mathew correlation coefficient (Mcc); see below. This measures can be computed as given below, such that TP, FN, TN and FP are the numbers of true positives, false negatives, true negatives and false positives, respectively.

$$Acc = 100 \times \frac{TP+TN}{TP+FN+TN+FP}$$
$$Se = 100 \times \frac{TP}{TP+FN}$$
$$Sp = 100 \times \frac{TP}{TN+FP}$$
$$Fm = 100 \times \frac{2 \times TP}{2 \times TP+FN+FP}$$
$$Mcc = 100 \times \frac{TP \times TN - FP \times FN}{\sqrt{((TP+FP) \times (TN+FN) \times (TP+FN) \times (TN+FP))}}$$

The first three measures are commonly used whereas Fm is prefered when a compromise between sensitivity and precision is desirable. Mcc measures the correlation between real and predicted classes and it is considered less biased towards class imbalance. We presented the predictive performances by the mean and the standard deviation (mean ± SD), over the 10 repetitions.

## 1.1 Features

Features used in this work are presented in Table 1, with references for the detailed descriptions. The prediction of the secondary structure in this work considered the energy model, as implemented in RNAfold [27] and UNAFold [28]. They predict the structure which gives the minimum free energy of folding (MFE). We kept the same parameters used in the original publications.

The sequence-structure features combined sequence nucleotide information and its predicted state at the secondary structure. In FS4, each feature represents the relative frequency of three contiguous nucleotides states at the secondary structure, fixing the middle character ($\{X_{sss}\}$, $X \in \{A, C, G, U\}$ and $s \in \{$paired, unpaired$\}$). Because the triplet-SVM script excludes sequences with multiple loops, we implemented a Python script to compute FS4 in any sequence. The motifs in FS5 give the counts of its occurrence in the sequence-structure string. This string is obtained by padding the nucleotide sequence with its respective predicted state at the secondary structure ({left-paired, right-paired, unpaired}). This set was obtained using the Python script provided in the authors' website. To compute FS1, we implemented a Python script, based on the microPred Perl pipeline. FS1 contains the largest diversity of features and depends on several independent scripts. We used the same scripts and options used in microPred. However, we used

RNAfold from ViennaRNA-2.0 [29] and UNAFold v3.8 [28], instead of the older versions used in microPred. Initially, we obtained implausible values for features based on pair-probabilities. They were linked to the function get_pr() from the RNA Perl package of ViennaRNA-2.0. We bypassed this problem by restarting Perl for each new sequence. FS2 and FS3 were obtained from FS1. A customized Python script computed the randfold (p) and the MFE with the RANDFold [34] package and merged these two features with FS4 to obtain FS6. FS7 was obtained merging FS2 and the seven additional features obtained using the Python script obtained from the authors' website.

## 1.2 Algorithms

The algorithms we adopted have different learning biases. This is important for the present work, since learning biases can play in favor of a feature set over others. SVM and RF are the two most applied algorithms for pre-miRNA classification, whereas $G^2DE$ offered class distribution interpretability. Similar interpretation would be obtained using RVKDE but [18] showed that RVKDE produced accuracies similar to $G^2DE$ and slightly lower than SVM, even though the number of kernels constructed by each algorithm were on average 920 (RVKDE), 361 (SVM) and six ($G^2DE$).

### 1.2.1 Support vector machines

SVMs deal with classification tasks by finding a hyperplane that separates training instances from two different classes with the maximum margin. The examples used to determine the hyperplane are the support vectors. Because many problems are not linearly separable, for these problems, the original feature space is mapped into a higher-dimensional space, where linear separation becomes feasible. Points from the original space are mapped to the new space by a kernel function. The RBF (radial basis function) kernel is a reasonable choice as it performs well for a wide range of problems [30]. For the training of SVMs, we used a Python interface for the library libsvm 3.12 [30]. This interface implements the C-SVM algorithm using the RBF kernel. The kernel parameters $\gamma$ and $C$ were tuned by 5-fold cross validation (CV) over the grid $2^{-5}, 2^{-3}, ..., 2^{15} \times 2^{-15}, 2^{-13}, ..., 2^{3}$. The pair $(C,\gamma)$ that led to the highest CV accuracy was used to train the SVMs using the whole training set. The induced model was then applied to the corresponding GEN test set.

*1.2.2 Random forest*

RF is an ensemble learning algorithm that induces a set of decision trees based on the concepts of "bagging" and random feature selection. Bagging is an important approach to estimate generalization error, whereas the latter is important to generate tree diversity. It was shown [31] that the strength of the ensemble depends on the strength of individual trees and the correlation between any two trees in the forest [32]. The number of features affects the strength of individual trees as well as the tree diversity, while the number of trees affects the generalization error. In order to obtain an ensemble with lower generalization error, a sufficiently large number of trees shall be chosen, taking into consideration two facts: RFs do not overfit, but limit the generalization error. This means that the number of trees has to be large enough to ensure lower generalization error, but after a certain value it does not have any effect on the generalization error estimate. For our experiments, we adopted the R package *randomForest* [32]. Each ensemble was generated over the grid (30, 40, 50, 60, 70, 80, 90, 100, 150, 250, 350, 450)×[ (0.5, 0.75, 1, 1.25, 1.5)*$\sqrt{d}$ ], representing respectively the number of trees and the number of features. The $\sqrt{d}$ is the default number of features tried in each node split and $d$ is the dimension of the feature space. We chose the ensemble with the lowest generalization error over the grid and applied it to the corresponding GEN test set.

*1.2.3 Generalized Gaussian density estimator*

$G^2DE$ [24] was designed to predict an instance class based on the probability density functions (pdf) of both positive and negative classes. Each pdf is fitted as mixture of generalized Gaussian components, using a limited user-defined number of components. One important feature of $G^2DE$ is to provide the coefficients and parameters associated with these generalized components [24].

The learning process of $G^2DE$ involves the estimation of the pdf parameters of each class, in addition to the weights of each component. If $k$ is the maximum number of components and the feature space has dimension $d$, the number of parameters will be $k(d+2)(d+1)/2$. An evolutionary optimization algorithm finds the solution by maximizing the number of instances correctly classified in the training set plus the likelihood of class distributions. It requires two user-defined parameters: the number of Gaussian components ($k$) and the number of individuals for the initial population in the genetic algorithm ($N$). The first was kept six as in [18], and $N$ was set to 100k, instead of 10k. High values of $N$ implicate in more running time. On the other hand, it is expected that high $N$ increases chances of findings an optimal solution. Since the solution is not deterministic, we ran $G^2DE$ five times and chose the solution which gave the highest CV accuracy. The number five was determined by us through computational experiments.

*1.2.4  Feature selection*

Once the prediction model is induced, the highest computational cost in the evaluation of putative pre-miRNAs is the feature extraction from the sequences to be classified. Since this procedure is performed on millions of sequences, we performed feature selection on different data sets, excluding features which depend on shuffled sequences, which have a higher extraction cost. We also analyzed the importance each one of the 85 features obtained by combining all features from FS1 and FS6, and 3 features from FS7. The analyses of feature importance was performed using the feature importance estimated by randomForest [32], and the F-score, as described in [33]. Briefly, randomForest estimates the importance a feature $x_i$ by computing the difference between the number of correctly classified out-of-bag (OOB) vectors before and after the permutation of $x_i$ in those vectors, during the training phase. For example, in the data used for this study, the number of OOB vectors is approximately 580. If an ensemble correctly classifies on average 500 and 36 OOB vectors before and after the permutation of $x_i$, the estimated importance is approximately 464. This value indicates that $x_i$ was crucial for the correct classification. However, the interpretation must consider that the importance is conditioned to the induced ensemble. Thus, another feature $x_j$ with importance 200 in the ensemble including $x_i$ could obtain an importance close to 464 in another ensemble excluding $x_i$. Nevertheless, this measure provides a createrion to evaluate the relevance of each feature given the whole set. Differently, the F-score estimates the ratio of between and within classes distances and is computed before the learning step. Features with higher F-scores are more likely to be more discriminative, even though there is no objective criterion to decide on a specific score cut-off. In our experiments, we trained SVMs eliminating features with F-score below different score thresholds.

## 2  Results and discussion
### 2.1  Effect of feature sets and training algorithm

Data dimensionality may affect the learning process, particularly for parametric models. In our experiments, G$^2$DE only converged to a predictive model for the feature set FS3, which has only seven features. This algorithm uses a genetic algorithm to estimate the parameters of Gaussian components and their corresponding weights. In order to obtain the individuals for the initial population, the genetic algorithm uses another algorithm which generates random covariance matrices. In our experiments, this algorithm generated non-positive definite matrices, which caused the non convergence of G$^2$DE for higher dimensions.

Table 2 shows that feature sets composed by a miscellany of RNA features produced higher classification performances than feature sets composed by sequence-structure patterns. However, the small differences in classification performance produced by FS3 and FS6, compared to FS1, FS2 and FS7, shows that either the diversity or the dimensionality may affect classification. Indeed, FS6 has a larger dimension than FS2, but it is not composed of the same level of feature diversity than FS2. On the other hand, FS3 is more diverse than FS6, but does not contain enough features to produce sensitivity comparable to FS2.

Nevertheless, it must be stressed that the highest predictive performances by the classifiers were not significantly different when the feature sets FS1, FS2, and FS7 were used (Table 2). As previously mentioned, FS2 and FS3 are subsets of FS1, while FS7 merged FS2 with seven additional features. These characteristics together with the very similar results obtained when using FS1, FS2, and FS7 suggest that the increase in the number of features leads to a limited increase of the predictive performance, even though the additional features were shown to be distinct features of pre-miRNAs [16, 20, 34, 35].

## 2.2 Feature discrimination and feature selection

Initially, we analyzed the importance of each one of the 85 features, obtained by merging FS1, FS6 and three features from FS7, when they were all used to induce classification ensembles by RF. In parallel, we also computed the F-scores. The Pearson correlation coefficient between the averages of these two measures was 75%, showing a relatively high correlation between importance and discrimination. Figure 1 shows the features whose importance was considered higher than 6. It can be seen that only 5 features obtained average importances higher than 40, corresponding to approx. 6% of the 580 OOB training instances. This results suggested that most of the 85 features may be redundant or irrelevant.

Interestingly, the features depeding on shuffled sequences appeared among those with the hightest importance or F-score. However, these features were not included in our feature selection step, due to their high computational cost and redundancy to the selected features. Moreover, since the features sets FS1 and FS7 share the 21 features of FS2 and they all produced classifiers with the highest predictive performances, we assumed that the relevant features for pre-miRNA classification were among these 21 common features. Therefore, the feature selection was performed using FS2, eliminating zD, which depends on shuffled sequences. The features selected (SELECT) from this set, in order of relevance, were: MFEI1, MFEI2, dG, dQ, dF, NEFE, Diff, dS, dS/L, |G-C|/L, |G-U|/L, %(G-U)/stems and MFEI3. Interestingly, six features of this set are energy-based measures (MFEI1, MFEI2, dG, NEFE, Diff, dS, MFEI3). The other relevant features are: entropy (dQ), compactness of the tree graph representation (dF), two thermodynamical features (dS and dS/L), normalized frequencies of G-C and G-U pairing (|G-C|/L, |G-U|/L, %(G-U)/stems).

The computational cost of each feature set was estimated by the computation time for a data set composed of 100 pre-miRNAs sequences randomly sampled from mirbase 20. Among the feature sets that produced the highest classification performance, FS1 had the highest cost (3:41h), followed by FS7 (1:18h), FS2 (1:17h) and FS6 (39:03min). Because SELECT and FS3 do not contain any stability measure, their costs are significantly lower than the cost for FS2. They took 2:17min and 10s to be computed. Among the sequence-structure based feature sets, FS5 took 3:24min to be computed, whereas FS4 took 2s.

The next comparison evaluates how the predictive performance associated with each feature set is affected by the use of different classifiers. For FS1, FS2, and FS7, the maximum difference in sensitivity and specificity between SVM and RF was 1.9% and 1.4%, respectively. The predictive performances of the three classifiers using FS3 were very similar. Thus, the learning biases of the three learning algorithms did not seem to have a significant effect on the predictive performance. This small effect of the learning biases is explained by the use of a sufficiently large training set, since most learning algorithms present a clear difference in their predictive performance only when small training sets are used.

### 2.3 Comparison with tools using the same algorithms and feature sets

In order to compare our results with tools used as references for our experiments, we predicted GEN test sets with those tools. Their main characteristics are summarized in Table 3. In Table 4 we show the predictive performance on the GEN sets obtained by the triplet-SVM, MiPred, microPred, $G^2DE$, mirident and HuntMi classifiers in our experiments. According to Table 4, except for the $G^2DE$ tool, which uses the $G^2DE$ algorithm, the predictive performance values obtained were much lower than the published values or the sensitivity was compromised by the specificity, or vice versa. The comparison in Table 4 used test sets obtained from mirbase 19, whereas the classifiers in those were induced with sequences of older releases. As the representativeness of the pre-miRNA population increases in newer releases, it is likely that the underlying distribution of the positive class would also changes. Therefore, the low sensitivities obtained by tools trained with old releases of mirbase, such as triplet-SVM, microPred, MiPred and mirident are not surprising. However, the low specificity values obtained by microPred and HuntMi, when compared to other older tools, were not expected. The specificity obtained by HuntMi in [20] was 72%, while we obtained 94% in our experiments, using the same algorithm and feature set. In contrast, the corresponding sensitivities were 99% and 88%. Likewise, the specificity of microPred [16] was 68%, while we found 95% using the same algorithm and feature set. Different results are usually obtained for experiments ran by different research groups, but not so different.

The largest loss in specificity is observed for microPred and HuntMi tools, which both correct for class imbalance. That is, they attempt to correct the bias due to training sets formed of imbalanced number of examples in each class. This imbalance may cause a bias towards the majority class in the learning algorithm. Ideally, the class imbalance correction would increase the sensitivity without dropping the specificity. MicroPred increased sensitivity from 83% to 90%, while the specificity dropped slightly from 99% to 97%. The imbalance rate, the ratio of positive to negative examples, in the data set was 1:13. Likewise, HuntMi reported sensitivity and specificity values of 94% and 95%, working under an imbalance rate of 1:57. Since the ideal imbalance rate is determined experimentally, it is plausible that the class imbalance correction methods applied by microPred and HuntMi caused generalization problems. A contributing effect, or alternative explanation, might be that HuntMi uses negative sequence which differ greatly from positive examples, whereas the negative sequences used in microPred and in our experiments were selected to be similar to positive sequences. The observed loss of specificity might be countered with modifications to the training procedures. However, the lack of generalization of microPred was also mentioned in [7, 36].

## 2.4 G+C content effect

G+C content is an important feature during the folding of hairpin-like RNA sequences. Because G+C-rich sequences have more alternative high-energy stable binding-pairs, the prediction of the corresponding secondary structure is more complex. We drew the slopes of sensitivity and specificity correspondents to 12.5%-G+C content quantiles, to have a picture of the predictive performance of the feature sets and the algorithms in predicting G+C-rich sequences. Figure 2 shows that the variation in specificity throughout the intervals is random. Nevertheless, the sensitivities depended on the feature set and on the algorithm. All feature sets dropped the sensitivities of RF classifiers in G+C-rich pre-miRNAs. However, when FS1, FS2, FS7 and SELECT (not shown) were used to train SVM classifiers, only random variations in sensitivity along the 12.5%-G+C content quantile intervals were obtained. These four feature sets have %G+C related features in common, such as MFEI1 and normalized frequencies of G-C and G-U pairing (|G-C|/L, |G-U|/L, %(G-U)/stems). As Figure 1 shows, except for MFEI1, the other features appear with relatively low importance in the induction of ensembles by RF. On the other hand, the support vectors from the SVM model contain all the features used. These results confirmed the importance of including %G+C related features to detect G+C-rich pre-miRNA.

## 2.5 Scope of the investigation

The research reported in this paper was carried out using human sequences, one of the species with the highest abundance of positive sequences. Assuming that the larger the amount of positive sequences, the larger the amount of information about the human pre-miRNA population, the investigation performed with human sequences allowed a more fair comparison of the features sets and learning algorithms. Nevertheless, as it has been indicated that the rising of novel miRNAs is highly correlated with morphological complexity [37–42], our results may vary for more distantly related species.

# 3 Conclusion

A considerable part of the computional cost involved in pre-miRNA prediction is due to the feature extraction from candidate sequences. Aiming to recommend effective and less costly sets of features, we investigated the discriminant power of seven RNA feature sets, under controlled sources of variation. Throughout extensive computational experiments, we showed that feature diversity is an important requirement in pre-miRNA recognition. Nevertheless, despite the discriminant power of individual features, higher dimensional sets did not produce higher classification performance classifiers. Based on these results, we proposed a smaller and less costly to compute subset of features, which produced classification performances as high as the produced by higher dimensional and more expensive sets. Because we attempted to avoid all possible sources of bias, we believe that the maximum classification performances reported here are the state-of-the-art for pre-miRNA prediction. Since these maximum classification performances are below experimentally feasible rates, other approaches to increase classification performance are welcome. As our tests showed, the tools used as references in our work either obtained low accuracies or the sensitivities or specificities were compromised.

# 4 Competing interests

The authors declare that they have no competing interests.

### Author's contributions

AS and AC conceived and supervised the study. IL assembled the data, implemented the scripts, ran the experiments and summarized the results. The three authors wrote and approved the final manuscript.

**Acknowledgements**

We thank Empresa Brasileira de Pesquisa Agropecuária (Embrapa Soybean) for the continuum financial support to the first author and Conselho Nacional de Desenvolvimento Científico e Tecnológico (CNPq) for the grant awarded to the first author between May 2010 and February 2012. We also thank Dr. Hsieh and Dr. Gudy for their valuable time taken to provide details of $G^2DE$ algorithm and HuntMi classifier, respectively.

## Figures

**Figure 1 - Average feature importance estimated during the induction of RF ensembles. Features with importance lower than five were omitted.**

The average feature importance drops-off quickly after the 10th feature, indicating that for each ensemble there are few distinguishing features.

**Figure 2 - Predictive performance of classifiers throughout $12.5\%$-quantile distribution of G+C content.**

The prediction of the secondary structure of $G + C$-rich sequences is more challenging. This figure shows that the classification of $G + C$-rich pre-miRNA sequences is also more complex. As the $G + C$ content increased, the sensitivity dropped, except when SVM was trained with feature sets including $\%G + C$-based features (FS1, FS2 and FS7).

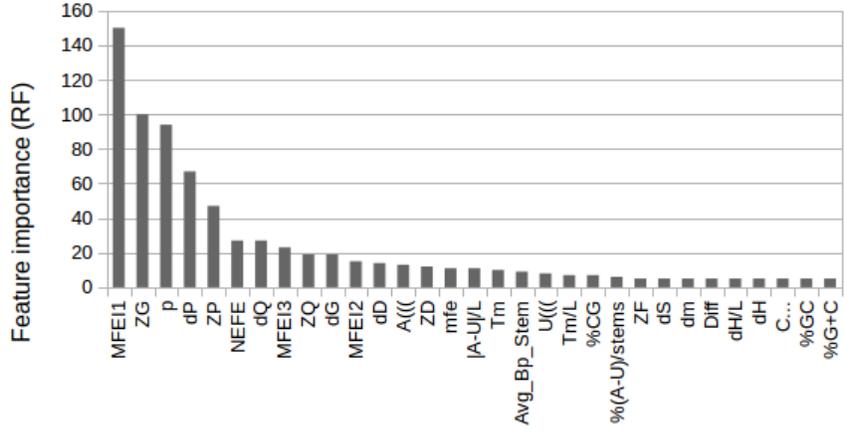

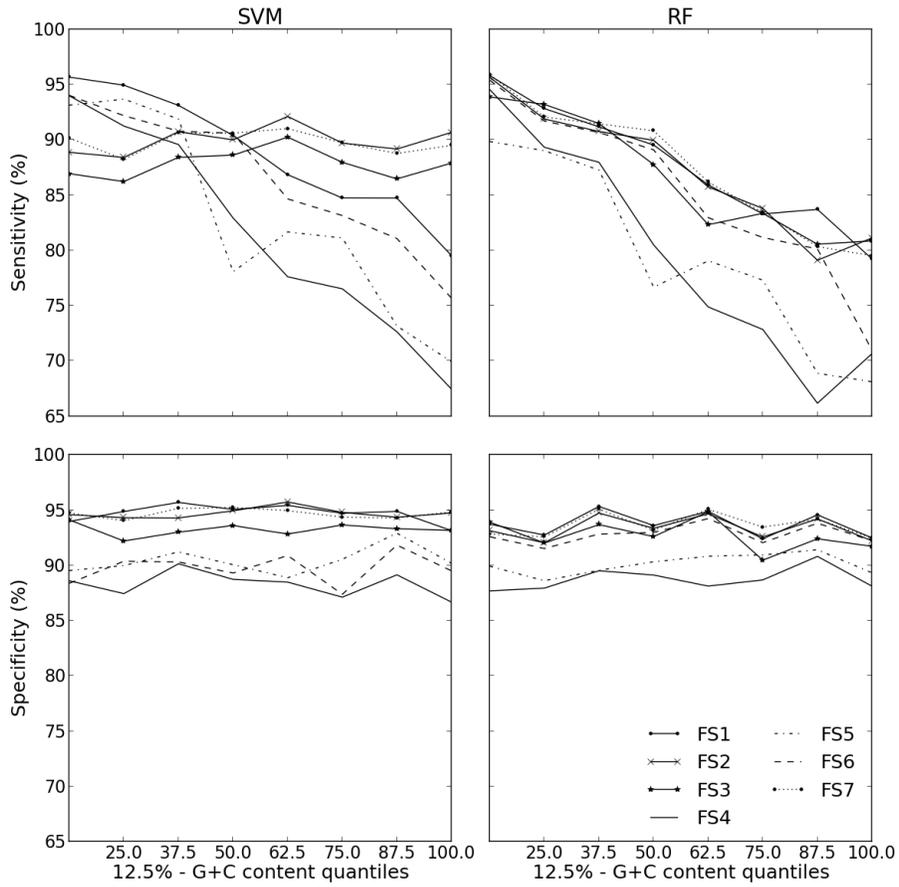

Table 1: Features used in each feature set. Detailed descriptions can be found in the corresponding references.

| FEATURE | REFERENCE | FS1 | FS2 | FS3 | FS4 | FS5 | FS6 | FS7 |
|---|---|---|---|---|---|---|---|---|
| Dinucleotide frequencies | [13] | x | | | | | | |
| $G+C$ content | [13, 35] | x | x | | | | | x |
| Maximal length of the amino acid string without stop codons | [20] | | | | | | | x |
| Low complexity regions detected in the sequence (%) | [20, 43] | | | | | | | x |
| Triplets | [10] | | | | x | | x | |
| Stacking triplets ($X_{(((}$, $X \in \{A, C, G, U\}$) | [10, 20] | | | | | | | x |
| Motifs ($ss$−substrings) | [23] | | | | | x | | |
| Minimum free energy of folding ($MFE$) | [27, 28] | | | | | | x | |
| Randfold ($p$) | [34] | | | | | | x | |
| Normalized MFE ($dG$) | [27] | x | x | x | | | | x |
| MFE index 1 ($MFEI_1$) | [35] | x | x | x | | | | x |
| MFE index 2 ($MFEI_2$) | [35] | x | x | x | | | | x |
| MFE index 3 ($MFEI_3$) | [16, 27] | x | x | | | | | x |
| MFE index 4 ($MFEI_4$) | [16, 27] | x | x | | | | | x |
| Normalized essemble free energy ($NEFE$) | [16, 27] | x | x | | | | | x |
| Normalized difference ($MFE - EFE$) ($Diff$) | [16, 27] | x | x | | | | | x |
| Frequency of the MFE structure ($Freq$) | [16, 27] | x | | | | | | |
| Normalized base-pairing propensity ($dP$) | [35, 44] | x | | x | | | | |
| Normalized Shannon entropy ($dQ$) | [35, 45] | x | x | x | | | | x |
| Structural diversity ($Diversity$) | [16, 35, 45] | x | x | | | | | x |
| Normalized base-pair distance ($dD$) | [16, 35, 45] | x | | x | | | | |
| Average base pairs per stem (Avg_Bp_Stem) | [16] | x | x | | | | | x |
| Average A-U pairs $|A-U|/L$ | [16] | x | x | | | | | x |
| Average G-C pairs $|G-C|/L$ | [16] | x | x | | | | | x |
| Average G-U pairs $|G-U|/L$ | [16] | x | x | | | | | x |
| Content of A-U pairs per stem %($A-U$)/$stems$ | [16] | x | x | | | | | x |
| Content of G-C pairs per stem %($G-C$)/$stems$ | [16] | x | x | | | | | x |
| Content of G-U pairs per stem %($G-U$)/$stems$ | [16] | x | x | | | | | x |
| Cumulative size of internal loops | [20] | | | | | | | x |
| Structure entropy ($dS$) | [16, 28, 35, 45] | x | x | | | | | x |
| Normalized structure entropy ($dS/L$) | [16, 28, 35, 45] | x | x | | | | | x |
| Structure enthalpy ($dH$) | [16, 28, 35, 45] | x | | | | | | |
| Normalized structure enthalpy ($dH/L$) | [16, 28, 35, 45] | x | | | | | | |
| Melting energy of the structure | [16, 28, 46] | x | | | | | | |
| Normalized melting energy of the structure | [16, 28, 46] | x | | | | | | |
| Topological descriptor (dF) | [35, 47] | x | x | x | | | | x |
| Normalized variants ($zG$, $zP$ and $zQ$) | [13, 35, 48] | x | | | | | | |
| Normalized variants ($zD$) | [13, 35, 48] | x | x | | | | | x |
| Normalized variants ($zF$) | [13, 35, 48] | x | | | | | | |

Table 2: Predicted accuracies (Acc), sensitivities (Se), specificities (Sp), F-measures (Fm) and Mathew Correlation Coefficients (Mcc) of classifiers trained with 1742 examples, presented as the mean and standard deviation (mean ± sd). Capital letters in columns indicate the performance cluster of each feature set, within algorithm (ALG). Lower case letters in columns indicate the cluster of each algorithms, within feature sets. Bold numbers represents the highest performances, which were not significantly different according to the clustering criteria in [49].

| ALG | FS | Acc | Se | Sp | Fm | Mcc |
|---|---|---|---|---|---|---|
| SVM | FS4 | E 85.6 ± 1.2 a | D 83.0 ± 1.9 a | D 88.4 ± 1.5 a | E 85.2 ± 1.3 a | E 71.4 ± 2.3 a |
|  | FS5 | D 87.4 ± 0.9 a | C 84.3 ± 1.5 a | C 90.5 ± 1.4 a | D 86.9 ± 0.9 a | D 74.9 ± 1.7 a |
|  | FS6 | C 89.8 ± 1.1 a | B 87.5 ± 1.5 a | C 93.0 ± 1.7 a | C 89.5 ± 1.1 a | C 79.8 ± 2.2 a |
|  | FS3 | B 90.6 ± 0.8 a | B 88.0 ± 1.3 a | B 93.3 ± 1.3 a | B 90.4 ± 0.9 a | B 81.4 ± 1.7 a |
|  | FS1 | **A 92.2 ± 0.9 a** | **A 89.7 ± 1.8 a** | **A 94.7 ± 0.8 a** | **A 92.0 ± 1.0 a** | **A 84.6 ± 1.8 a** |
|  | FS2 | **A 92.4 ± 0.9 a** | **A 90.1 ± 1.6 a** | **A 94.7 ± 0.6 a** | **A 92.2 ± 1.0 a** | **A 84.9 ± 1.8 a** |
|  | FS7 | **A 92.3 ± 1.0 a** | **A 89.9 ± 1.1 a** | **A 94.7 ± 0.9 a** | **A 92.1 ± 0.9 a** | **A 84.7 ± 1.6 a** |
|  | SELECT | **A 92.3 ± 0.9 a** | **A 90.0 ± 1.3 a** | **A 94.6 ± 1.0 a** | **A 92.1 ± 0.9 a** | **A 84.6 ± 1.7 a** |
| RF | FS4 | E 84.8 ± 1.1 b | D 81.2 ± 1.8 b | C 88.3 ± 1.3 a | E 84.2 ± 1.2 b | E 69.8 ± 2.1 b |
|  | FS5 | D 85.7 ± 0.7 b | D 81.2 ± 0.8 b | B 90.3 ± 1.4 a | D 85.1 ± 0.6 b | D 71.8 ± 1.5 b |
|  | FS6 | C 88.7 ± 1.4 b | C 86.6 ± 1.5 b | A 89.8 ± 1.6 b | C 88.5 ± 1.4 b | C 77.4 ± 2.8 b |
|  | FS3 | C 90.0 ± 1.0 b | C 86.9 ± 1.4 b | A 93.0 ± 1.1 a | C 89.6 ± 1.0 b | C 80.1 ± 1.9 b |
|  | FS1 | A 91.5 ± 1.0 b | A 89.1 ± 1.1 a | A 93.9 ± 1.2 a | A 91.3 ± 1.0 b | A 83.1 ± 1.9 b |
|  | FS2 | A 90.9 ± 1.0 b | B 88.1 ± 1.2 b | A 93.8 ± 1.3 b | A 90.7 ± 1.1 b | A 82.0 ± 2.1 b |
|  | FS7 | A 91.1 ± 0.8 b | B 88.5 ± 1.3 b | A 93.7 ± 1.3 b | A 90.9 ± 1.0 b | A 82.3 ± 2.0 b |
|  | SELECT | B 90.5 ± 0.9 b | C 87.4 ± 1.0 b | A 93.6 ± 1.4 b | B 90.2 ± 0.9 b | B 81.2 ± 1.9 b |
| G$^2$DE | FS3 | 90.2 ± 0.9 | 87.4 ± 1.5 | 93.1 ± 0.9 | 89.9 ± 0.9 | 80.6 ± 1.8 |

**Tables**
**Table 1** - **Features used in each feature set. Detailed descriptions can be found in the corresponding references.**
**Table 2** - **Predicted performances of classifiers trained with 1742 examples, presented as the mean and standard deviation ($Mean \pm SD$). Acc=accuracy; Se=sensitivity; Sp=specificity; Fm=F-measure; Mcc=Mathew Correlation Coefficient. Bold numbers represents the highest classification performances, which were not significantly different according to the clustering criteria in [49].**
**Table 3** - **Main characteristics of tools used as references in this work. BP=*Number of base pairs on the stem*, MFE=*Minimum Free Energy of the secondary structure*, noML=*no Multiple Loops*, RR=*Removed Redundancies*, E-value$\leq 10^2$=expected value in BLASTN against mirbase, ExpVal=Only experimentally validated precursors and RF=*Random forest***
**Table 4** - **Predicted performance of tools used as references in our GEN experiments test sets. Results presented as the mean and the standard deviation ($Mean \pm SD$). Acc=accuracy; Se=sensitivity; Sp=specificity; Fm=F-measure; Mcc=Mathew Correlation Coefficient.**

Table 3: Main characteristics of tools used as references in this work. BP=*Number of base pairs on the stem*, MFE=*Minimum Free Energy of the secondary structure*, noML=*no Multiple Loops*, RR=*Removed Redundancies*, E-value≤ $10^2$=expected value in BLASTN against mirbase, ExpVal=Only experimentally validated precursors and RF=*Random forest*

| TOOL | ALGORITHM | #FEATURES | PRE-PROCESSING | | TRAIN | | SOURCE | |
|---|---|---|---|---|---|---|---|---|
| | | | (+) | (−) | (+) | (−) | (+) | (−) |
| Triplet−SVM | SVM | 32 | noML | $BP > 18$ $MFE < -15$ noML | 163 | 168 | 5.0 | CDS |
| MiPred | RF | 34 | noML | $BP > 18$ $MFE < -15$ $50 < len < 138$ noML | 163 | 168 | 8.2 | CDS |
| microPred | SVM | 21 | RR | $len < 151$ RR | Not given clearly | Not given clearly | 12 | CDS ncRNAs |
| $G^2DE$ | $G^2DE$ | 7 | RR noML | $BP > 18$ $MFE < -25$ noML | 460 | 460 | 12.0 | CDS |
| mirident | SVM | 1300 | RR noML | $BP > 18$ $MFE < -25$ $50 < len < 138$ RR noML | 484 | 484 | 11.0 | CDS |
| HuntMi | RF | 28 | ExpVal | $E - value \leq 10^2$ | Not given clearly | Not given clearly | 17.0 | CDS mRNA ncRNA |

Table 4: Predicted performance of tools used as references in our GEN experiments test sets. Results are presented as the mean and the standard deviation ($Mean \pm SD$). Acc=accuracy; Se=sensitivity; Sp=specificity; Fm=F-measure; Mcc=Mathew Correlation Coefficient.

| TOOL | Acc | Se | Sp | Fm | Mcc |
|---|---|---|---|---|---|
| Triple-SVM | 78.8 ± 1.3 | 64.7 ± 2.1 | 92.9 ± 1.3 | 75.3 ± 1.7 | 60.1 ± 2.5 |
| MiPred | 86.8 ± 0.9 | 76.8 ± 1.6 | 96.8 ± 0.9 | 85.3 ± 1.1 | 75.1 ± 1.7 |
| microPred | 69.9 ± 1.7 | 72.1 ± 1.7 | 67.6 ± 2.7 | 70.6 ± 1.5 | 39.8 ± 3.3 |
| $G^2DE$ | 90.6 ± 0.9 | 89.2 ± 1.2 | 93.3 ± 1.6 | 90.5 ± 0.9 | 81.4 ± 1.8 |
| Mirident | 85.5 ± 1.0 | 88.2 ± 1.1 | 82.9 ± 1.2 | 85.9 ± 1.0 | 71.2 ± 2.1 |
| HuntMi | 85.1 ± 2.1 | 98.7 ± 0.8 | 71.6 ± 4.2 | 86.9 ± 1.6 | 73.0 ± 3.5 |